\documentclass[showpacs,preprintnumbers,
amsmath,amssymb,groupedaddress,superscriptaddress]{revtex4}
\usepackage{graphicx}
\usepackage{dcolumn}
\usepackage{bm}
\usepackage{epstopdf}
\usepackage{hyperref, bm, slashed, array, mathtools, amsfonts, graphicx}
\begin{document}
\title{\hskip11cm NT@UW-14-25\\Charge Symmetry Breaking in Electromagnetic Nucleon Form Factors in Elastic Parity-Violating Electron-Nucleus  Scattering}

\author{  Gerald A. Miller}
\address{Department of Physics, University of Washington,  Seattle, WA 98195-1560 
}

\date{\today}

\begin{abstract}
The effects of  charge symmetry breaking in nucleon electromagnetic form factors on
   parity-violating elastic
  electron-$^{12}$C scattering is studied, and found to be much smaller than other known 
  effects.  The analysis of a planned experiment is discussed.

\end{abstract}


\maketitle

\section{Introduction}

The  Standard
Model can be tested in  low energy electron-nucleus scattering~\cite{Fei75,Wal77,
Don79,Mus94}. For nuclei with $J^\pi=0^+$  the PV
asymmetry acquires a very simple, model-independent expression in
terms of the weak  nuclear charge, with nuclear structure effects
canceling out if  
the nuclear ground state is purely isospin 0, and if effects of strangeness and
charge symmetry breaking in the nucleon electromagnetic form factors can be ignored.

Indeed, plans are underway to measure the weak charge of the $^{12}C$ nucleus as part of the P2 experiment at Mainz~\cite{Gerz:2013ema}. This is a low-momentum transfer PV elastic electron-nucleus scattering experiment
with the 
 aim of reaching a  relative precision of 0.3\%.
Much work has already been done on the effects of nuclear  isospin mixing~\cite{Moreno:2008ve} as well as nucleon strangeness~\cite{Mus94,GonzalezJimenez:2011fq,Gonzalez-Jimenez:2014bia}.
Our purpose here is to asses the effects of charge symmetry breaking (CSB)  of nucleon electromagnetic form factors on nuclear parity-vioalting electron scattering.

In the  following we first discuss the general expression for the PV asymmetry, including all correction terms expected to be relevant. Our focus is on comparing the computed size of the
 CSB effects with the strangeness and nuclear isospin violation effects that  are already  in the literature for the kinematics $0\le Q^2 \le 0.063 $ GeV$^2$ of the planned P2 experiment~\cite{Gerz:2013ema}.

\section{Parity-violation asymmetry}
Polarized    electron  elastic scattering from unpolarized nuclei has 
been used to study parity violation, because  both electromagnetic (EM) and
weak interactions contribute to the process via $\gamma$ and $Z^0$
exchange.  The PV asymmetry is given by~\cite{Mus94}
\begin{equation}\label{asymmetry_sigmas}
{\mathcal A}=\frac{d\sigma^+ - d\sigma^-}{d\sigma^+ + d\sigma^-}\, ,
\end{equation}
where $d\sigma^+ (d\sigma^-)$ is the cross section for electrons
longitudinally polarized parallel (antiparallel) to their
momentum. The asymmetry $\mathcal A$ for a target state of $J^\pi
=0^+$,
predicted by 
the Standard Model can be written as
\begin{equation}
{\mathcal A}= \frac{G_F}{2\pi\alpha\sqrt{2}}Q^2a_A\frac{ \tilde{F}_{C0} (q)}{ F_{C0}(q)},\\
\end{equation}
where $G_F$ and $\alpha$ are the Fermi and fine-structure coupling
constants,  $Q^2$ is the negative of the square of the
four-momentum transfer in the
scattering process, 
 $a_A=-1$, and the terms $F_{C0}$ and $ \tilde{F}_{C0}$  are the electromagnetic and weak neutral current 
 nuclear form factors.  This result is obtained   in the Plane Wave Born Approximation
by keeping only the square of the photon-exchange amplitude for the spin-averaged EM cross section and using the interference between the $\gamma$ and $Z^0$ exchange  amplitudes in the cross section difference.

For  $N=Z$ nuclear ground states  that are pure     isospin zero,  only
isoscalar matrix elements contribute and the weak and EM form factors
obey the proportionality relation:
\begin{equation}
\tilde{F}_{C0} (q) = \beta ^{(0)}_V F_{C0}(q),\, 
\end{equation}
so that  the resulting PV asymmetry, $ {\mathcal A}^0$   depends only on fundamental constants:
\begin{equation}
 {\mathcal A}^0 \equiv \left[ \frac{G_F\,Q^2 \,
}{2\pi \alpha \sqrt{2}} \right] a_A \beta ^{(0)}_V \cong 3.22 \times
10^{-6} {Q^2\over \text{ fm}^{-2}} 
\label{referencevalue}
\end{equation}
where, within the Standard Model, $a_A \beta_V^{(0)}=2\sin^2\theta_W$, with 
$\theta_W$ as the weak mixing angle.  This  proportionality with 
$\sin^2\theta_W$, provides an ability to test the Standard Model, which has intrigued many.
 But one must handle corrections which occur as the result of 
 the effects of nuclear isospin
mixing, strangeness content and charge symmetry breaking in nucleon electromagnetic form factors.

\newcommand{\eq}[1]{Eq.~(\ref{#1})}

\newcommand{\hook}{\mathbin{\rule[-.2ex]{.7em}{.03em}\rule[-.2ex]{.03em}{1.4ex}}}
\let\vaccent=\v 
\renewcommand{\v}[1]{\ensuremath{\mathbf{#1}}} 
\newcommand{\gv}[1]{\ensuremath{\mbox{\boldmath$ #1 $}}} 
\newcommand{\uv}[1]{\ensuremath{\mathbf{\hat{#1}}}} 
\newcommand{\abs}[1]{\left| #1 \right|} 
\newcommand{\avg}[1]{\left< #1 \right>} 
\let\underdot=\d 
\renewcommand{\d}[2]{\frac{d #1}{d #2}} 
\newcommand{\dd}[2]{\frac{d^2 #1}{d #2^2}} 
\newcommand{\p}{\partial}
\newcommand{\pd}[2]{\frac{\partial #1}{\partial #2}} 
\newcommand{\pdd}[2]{\frac{\partial^2 #1}{\partial #2^2}} 
\newcommand{\pdc}[3]{\left( \frac{\partial #1}{\partial #2}
 \right)_{#3}} 
\newcommand{\fd}[2]{\frac{\delta #1}{\delta #2}} 
\newcommand{\ifd}[1]{\frac{\delta}{i\delta #1}} 
\newcommand{\ifdad}[1]{\frac{i\delta}{\delta #1}} 
\newcommand{\ifdp}[1]{\frac{(2\pi)^4\delta}{i\delta #1}} 
\newcommand{\ifdadp}[1]{\frac{i(2\pi)^4\delta}{\delta #1}} 
\newcommand{\ket}[1]{\left| #1 \right>} 
\newcommand{\bra}[1]{\left< #1 \right|} 
\newcommand{\braket}[2]{\left< #1 \vphantom{#2} \right|
 \left. #2 \vphantom{#1} \right>} 
\newcommand{\mbraket}[3]{\left< #1 \vphantom{#2#3} \right|
 #2 \left| #3 \vphantom{#1#2} \right>} 
\newcommand{\grad}[1]{\gv{\nabla} #1} 
\let\divsymb=\div 
\renewcommand{\div}[1]{\gv{\nabla} \cdot #1} 
\newcommand{\curl}[1]{\gv{\nabla} \times #1} 
\newcommand{\tr}{\text{Tr}} 
\let\bar=\smallbar 
\newcommand{\bar}[1]{\overline{#1}} 
\newcommand{\fs}[1]{\slashed{#1}} 
\let\tilde=\widetilde
\newcommand{\bea}{\begin{eqnarray}}
\newcommand{\eea}{\end{eqnarray}}

\bigskip

We begin to assess these different effects, starting by taking matrix elements of the basic weak interaction.
In the Standard Model the  weak  neutral vector coupling  between a $Z$-boson and a quark  is given by  ${1\over2}(\tau^3 - 4 s_W Q_q)$, where $s_W=\sin^2\theta_W$ and $Q_q$ is the quark charge in units of the proton charge. We shall use $s_W=0.234$ for our numerical work. Then the nucleon ($N$) weak form factors are given in terms of the quark and electromagnetic current form factors as
\begin{equation}
     F_{1,2}^{Z,N} = {1\over2}\left( F_{1,2}^{u,N} -  F_{1,2}^{d,N} -  F_{1,2}^{s,N} - 4s_W  F_{1,2}^{em,N}\right),
\end{equation}
where $F_{1,2}^q$ is the contribution of the quark ($q$) to the nucleon Dirac or Pauli form factor. 

The CSB form factors $F^{\fs{s}}$ and $F^{\fs{v}}$ are  related to matrix elements of an isoscalar current $j_s^\mu = \frac{1}{6}(\bar{u}\gamma^\mu u + \bar{d}\gamma^\mu d)$ and isovector current $j_v^{3,\mu} = \frac{1}{2}(\bar{u}\gamma^\mu u - \bar{d}\gamma^\mu d)\tau^3$ by
\begin{eqnarray}
    \bar{u}_N(P+q)\left[ F_1^\fs{s}(Q^2)\gamma^\mu + F_2^{\fs{s}}(Q^2)\frac{i\sigma^{\mu\nu}q_\nu}{2m_N} \right]u_N(P) &=& \mbraket{p}{j^\mu_s}{p} - \mbraket{n}{j^\mu_s}{n},\\\nonumber
    \bar{u}_N(P+q)\left[ F_1^\fs{v}(Q^2)\gamma^\mu + F_2^{\fs{v}}(Q^2)\frac{i\sigma^{\mu\nu}q_\nu}{2m_N} \right]u_N(P) &=& \mbraket{p}{j^\mu_v}{p} + \mbraket{n}{j^\mu_v}{n},
\end{eqnarray}
with $m_N$ as the average nucleon mass.
We can then express isoscalar and isovector combinations as
\begin{eqnarray}
     F_{1,2}^{Z,p} +  F_{1,2}^{Z,n}  
    &=&  F_{1,2}^{\fs{v}} -  F_{1,2}^s - 2s_W( F_{1,2}^{em,p} +  F_{1,2}^{em,n}),\label{gzp}\eea
     where \bea F_{1,2}^{\fs{v}} \equiv {1\over2}(F_{1,2}^{u,p}-F_{1,2}^{d,p}+F_{1,2}^{u,n}-F_{1,2}^{d,n}),\eea
     and  
    \bea F_{1,2}^{Z,p} -  F_{1,2}^{Z,n} 
    &=& 
 ((1 - 2s_W)\,( F_{1,2}^{em,p} -  F_{1,2}^{em,n}) -  F_{1,2}^{\fs{s}},\label{gzm}
\end{eqnarray}
where
\bea F_{1,2}^{\fs{s}}\equiv {1\over 6}(F_{1,2}^{u,p}+F_{1,2}^{d,p}-F_{1,2}^{u,n}-F_{1,2}^{d,n}).\eea

These form factors are multiplied by the point-nucleon form factors $F_{p,n}(Q^2)$ of the nucleus to obtain the 
form factors $F_{C0}$ and $\tilde{F}_{C0}$. This assumes that all of the nuclear strangeness lies within individual nucleons. Any other nuclear strangeness would arise from  an $s$ quark confined to one baryon and an $\bar{s}$ confined to another nucleon.   The existence of such exotic components is highly suppressed by large energy denominators and is ignored here. We also neglect meson exchange currents, as these are expected to be very small~\cite{Moreno:2008ve}.

 The relevant ratio $\frac{\tilde{F}_{C0} (Q^2)}{F_{C0}(Q^2)}$ is given by
 \bea \frac{\tilde{F}_{C0} (Q^2)}{F_{C0}(Q^2)}={G_E^{Z,p}(Q^2)F_p(Q^2)+G_E^{Z,n}(Q^2)F_n(Q^2)\over
 G_E^{em,p}(Q^2)F_p(Q^2)+G_E^{em,n}(Q^2)F_n(Q^2)}
 ,\label{frat}\eea
 where $G_E^{Z,N},\,G_E^{em,N}$ are the Sach's   electric form factors computed using the average value of the nucleon mass.   The above expression is obtained neglecting the leading  term in the nucleon current, a term of  the order of the nucleon momentum divided by the nucleon mass.  The equations in Ref.~\cite{Moreno:2008ve} show for a $C^{12}$ nucleus,  such terms are at most approximately $Q^2/(12m_N^2)\approx 5\times 10^{-3} $ of the small correction terms we keep  at the low values of momentum transfer  of interest to the experiment~\cite{Gerz:2013ema}.

 Next we simplify \eq{frat} by defining
 \bea
 F_p(Q^2)\equiv \bar{F}(Q^2)+{1\over2}\Delta F(Q^2)\\
 F_n(Q^2)\equiv \bar{F}(Q^2)-{1\over2}\Delta F(Q^2)\\
 G_\pm^Z(Q^2)\equiv G_E^{Z,p}(Q^2)\pm G_E^{Z,n}(Q^2)\\
  G_\pm^{em}(Q^2)\equiv G_E^{em,p}(Q^2)\pm G_E^{em,n}(Q^2).
\eea
 Using this notation  
 and  keeping the leading term and those of first-order in the corrections $G_E^s,G_E^{\fs{v}}$ and $\Delta F$ gives 
  \bea& \frac{\tilde{F}_{C0} (Q^2)}{F_{C0}(Q^2)} 
= -2s_W+{G_E^{\fs{v}}-G_E^s\over G_+^{em}}+{(1-2s_W)^2G_-^{em}\over G_+^{em}}{\Delta F\over 2\bar{F}}. 
 \eea
The net result is that
\bea
 {\mathcal A}=\left[ \frac{G_F\,Q^2 \,
}{2\pi \alpha \sqrt{2}} \right] \left  (2s_W-{G_E^{\fs{v}}-G_E^s\over G_+^{em}}-{(1-2s_W)^2G_-^{em}\over G_+^{em}}{\Delta F\over 2\bar{F}}\right).
 \eea
The nucleon electromagnetic form factors of Kelly\cite{Kelly:2004hm} are used in our calculations.

 One may define the correction to the   $2s_W$  term as $C(Q^2)\equiv-{G_E^{\fs{v}}-G_E^s\over G_+^{em}}-{(1-2s_W)^2G_-^{em}\over G_+^{em}}{\Delta F\over 2\bar{F}}$  so that
 \bea
 {\mathcal A}=\left[ \frac{G_F\,Q^2 \,
}{2\pi \alpha \sqrt{2}} \right] \left  (2s_W+C(Q^2)\right).\label{cdef}
 \eea
One way to analyze an  experiment is to  make an extrapolation  linear in $Q^2$ to determine the value of
 $s_W$, so we shall be concerned with the linearity of $C(Q^2)$.
 
 \section{The Correction  Term $C(Q^2)$}
 
 We consider the three contributions to $C(Q^2)$.
 \subsection{Charge symmetry breaking (CSB) of the electromagnetic form factors}
 We have previously evaluated~\cite{Wagman:2014nfa} the leading-order  CSB effects of the pion cloud of the nucleon 
and of vector mesons which contribute to the leading low energy constant~\cite{Kubis:2006cy}.
Our previous work did not obtain the separate terms $F_{1,2}^{\fs{v},\fs{s}}.$ This is done here. The pionic terms are given by
\begin{eqnarray}
    F_1^{\fs{s}} =  -\left( \frac{g_Am_N}{f_\pi} \right)^2\left[ \tilde{I}_1(Q^2,m_p,m_n) - \tilde{I}_1(Q^2,m_n,m_p) \right],
\end{eqnarray}
\begin{eqnarray}
    F_2^{\fs{s}} =  2\left( \frac{g_Am_N}{f_\pi} \right)^2\left[ I_2(Q^2,m_p,m_n) - I_2(Q^2,m_n,m_p) \right],
\end{eqnarray}
\begin{eqnarray}
    F_1^{\fs{v}} &=&   \left( \frac{g_Am_N}{f_\pi} \right)^2\left[ \tilde{I}_1(Q^2,m_p,m_n) - \tilde{I}_1(Q^2,m_n,m_p) - \tilde{J}_1(Q^2,m_p,m_n) + \tilde{J}_1(Q^2,m_n,m_p) \right],\\ 
    F_2^{\fs{v}}  
    &=&  \left( \frac{g_Am_N}{f_\pi} \right)^2\left[ -2I_2(Q^2,m_p,m_n) + 2I_2(Q^2,m_n,m_p) - 2J_2(Q^2,m_p,m_n) + 2J_2(Q^2,m_n,m_p) \right].
\end{eqnarray}
The values of the  axial vector coupling constant, $g_A$, the pion decay constant $f_\pi$, and the average nucleon mass are presented in Ref.~\cite{Wagman:2014nfa}. The terms $ \tilde{I}_1(Q^2,m_p,m_n),\, I_2(Q^2,m_p,m_n),\,\tilde{J}_1(Q^2,m_n,m_p)$, and $J_2(Q^2,m_p,m_n)$  
are obtained from the relevant Feynman diagrams and 
are specified in Eqs(9) and (10) of
Ref.~\cite{Wagman:2014nfa}.

We also need to include our resonance saturation assumptions for the phenomenologically unconstrained contact terms $\kappa^{\fs{s}}$ and $\kappa^{\fs{v}}$ discussed in Ref.~\cite{Wagman:2014nfa}.  These terms dominate the CSB contribution to $G_E$ of the proton~\cite{Kubis:2006cy}.  The $\omega$ couples to isoscalar currents, and so the diagram $\omega \rightarrow \rho$ where the $\omega$ couples to a current and then mixes with a $\rho$ that couples to a nucleon as an isovector contributes to $F^{\fs{s}}$. Conversely the $\rho$ couples to isovector currents, so the diagram with $\rho\rightarrow\omega$ contributes to $F^{\fs{v}}$. This gives
\begin{eqnarray}
    F_1^{VM,\fs{s}} &=& g_\rho  F_\omega \Theta_{\rho\omega} \frac{Q^2}{m_V(m_V^2 + Q^2)^2},\hspace{25pt} F_2^{VM,\fs{s}} = -g_\rho \kappa_\rho F_\omega \Theta_{\rho\omega} \frac{m_V}{(m_V^2 + Q^2)^2},\\\nonumber
    F_1^{VM,\fs{v}} &=& g_\omega  F_\rho \Theta_{\rho\omega} \frac{Q^2}{m_V(m_V^2+Q^2)^2},\hspace{25pt} F_2^{VM,\fs{v}} = -g_\omega \kappa_\omega F_\rho \Theta_{\rho\omega} \frac{m_V}{(m_V^2+Q^2)^2}.
\end{eqnarray}

The effects of CSB are to be compared with those of strangeness in the nucleon.

 \subsection{Strangeness}
The effects of strangeness on nucleon electromagnetic form factors has been  parameterized~\cite{Moreno:2008ve} as 
  \begin{equation}
G_{E}^{(s)} = \rho_s \tau G_D^V \xi_E^{(s)}, \qquad G_{M}^{(s)} =
\mu_s G_D^V \, ,
\label{str_para_1}
\end{equation}
with  (for instance, \cite{Mus94})  
\begin{equation}
G_D^V = (1+4.97 \tau)^{-2}, \quad \xi_E^{(s)}=(1+ 5.6 \tau)^{-1} \, .
\label{str_para_2}
\end{equation}
The parameter $\rho_s$ and $\mu_s$ are constrained by PV electron
scattering measurements on hydrogen, deuterium and helium-4. Ref.~\cite{Moreno:2008ve} used the range
$-1.5<\rho_s <1.5$. Later work~\cite{Gonzalez-Jimenez:2014bia}  made a statistical analysis of the full set of parity-violating asymmetry data for elastic electron scattering. This found $\rho_s=0.92\pm0.58$. We use this range of values in our  numerical work.  However, experiments on deep inelastic scattering restrict the
$s$ and $\bar{s}$ parton distribution functions to very small values~\cite{Bentz:2009yy}
and  reality may correspond to an order of magnitude  smaller 
values of $\rho_s$~\cite{us}.

\subsection{Nuclear Isospin violation}
Ref.~\cite{Moreno:2008ve} used a Skyrme type density-dependent interaction to generate the ground state wave function in the Hartree-Fock plus BCS approximation. This procedure yields  ground state densities for $^{12}$C,$^{24}$Mg, $^{28}$Si and $^{32}$S nuclei which give  computed nuclear charge form factors in excellent agreement with electron scattering data.  The difference in proton and neutron charge densities is generated mainly by the Coulomb interaction. 
 Here we use the result of a different formalism: a   new nuclear density functional of Bulgac {\it et al.}~\cite{ab}. This calculation produces nuclear densities constrained by nuclear binding energies  and  charge densities for the entire periodic table. For  $^{12}$C the calculation of ~\cite{ab} causes almost exactly the same effects in the  PV asymmetry as the one of Ref.~\cite{Moreno:2008ve}. This lends credence to the idea that the many-body nuclear theory is under control. Its uncertainties would not impact experimental extractions of
the weak mixing angle or strangeness content.  Furthermore, the effects of isospin-violating strong forces are much, much smaller than  those of the Coulomb interaction for all 
nuclei~\cite{AUERBACH:1972,HM79,Miller:1990iz}.

\section{Results and Conclusions}
Our aim is to present calculations relevant for the planned experiment~\cite{Gerz:2013ema}. Therefore the momentum transfer range is restricted to $0\le Q^2\le 0.0625$ GeV$^2$.

\begin{figure}[t]
\includegraphics[width=6.5cm,height=6cm]{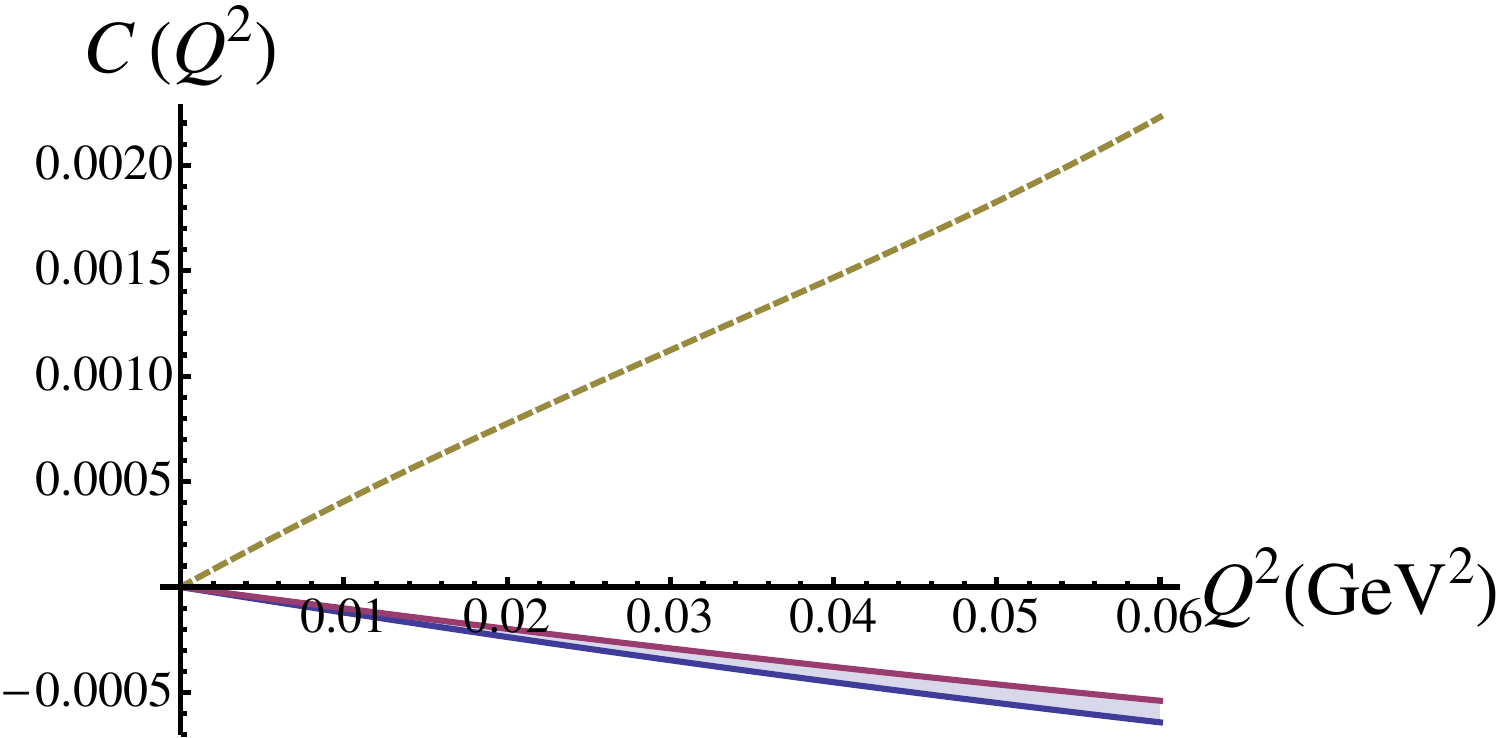}
\caption{(color online)Contributions to the correction $C(Q^2)$ due to nuclear isospin violation (dashed) and CSB in the nucleon form factors (solid) for two sets of meson-nucleon coupling constants, see text. }\label{NIVCSB}\end{figure}

We begin  by comparing the effects of charge symmetry breaking (CSB) in nucleon electromagnetic form factors with the
effects of nuclear isospin violation, see Fig.~\ref{NIVCSB}. As expected, the nuclear effects are far larger than
those of the nucleon. The range of curves for the  CSB terms is obtained from using the compilations of  Refs.~\cite{Dumbrajs:1983jd} and \cite{Ericson:1988gk}.  If the value of $Q^2$ were increased by about 15\%, the effects of nuclear isospin would become very large an non-linear in the variable $Q^2$. This feature is in agreement with the results of Ref.~\cite{Moreno:2008ve}. However, the restriction of the value of $Q^2$ to an upper limit of $0.0625$ GeV$^2$ is sufficient to ensure a linear behavior. Note also that  any effects of the uncertainty in the  nuclear isospin violation terms (expected to be no more than 5\%)
are expected to be far smaller than the uncertainty goal of the planned experiment~\cite{Gerz:2013ema}.

\begin{figure}[h]
\includegraphics[width=6.5cm,height=6cm]{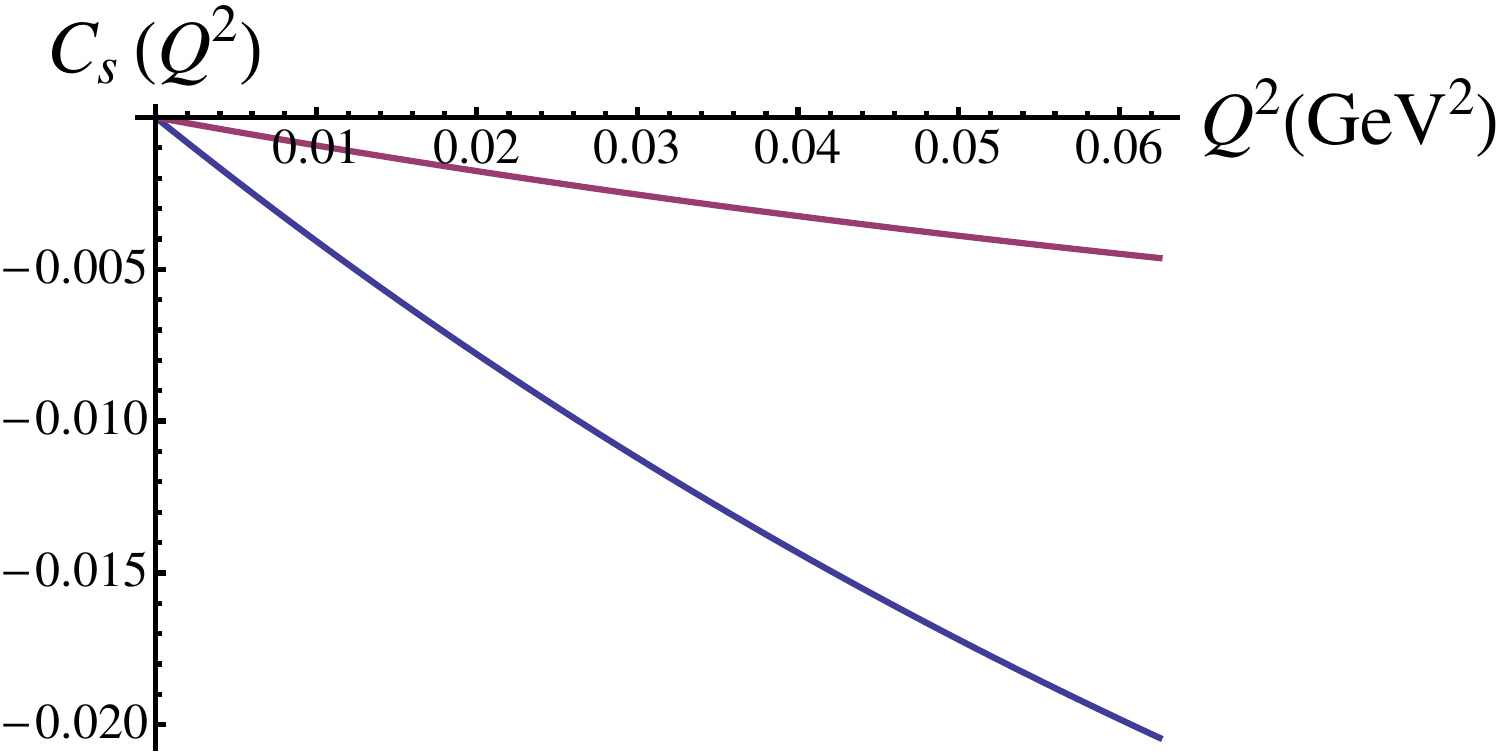}
\caption{(color online)Contributions to the correction $C(Q^2)$ due to strangeness. The two curves are obtained using the upper (+0.15) and lower (0.34) limits on $\rho_s$ from~\cite   {Gonzalez-Jimenez:2014bia}. }\label{Strange}\end{figure}

Next we assess the effects of nucleon strangeness using the range of values of from~\cite{Gonzalez-Jimenez:2014bia}, $\rho_s=0.92\pm0.58$, see Fig.~\ref{Strange}. Comparing Figs.~~\ref{NIVCSB} and ~\ref{Strange}
shows that the CSB effects are generally more than an order of magnitude smaller than those of nucleon strangeness obtained from these limits.   This statement is consistent with that of Ref.~\cite{Wagman:2014nfa}, which compared proton CSB effects with experimental uncertainties. 

\begin{figure}[t]
\includegraphics[width=6.5cm,height=6cm]{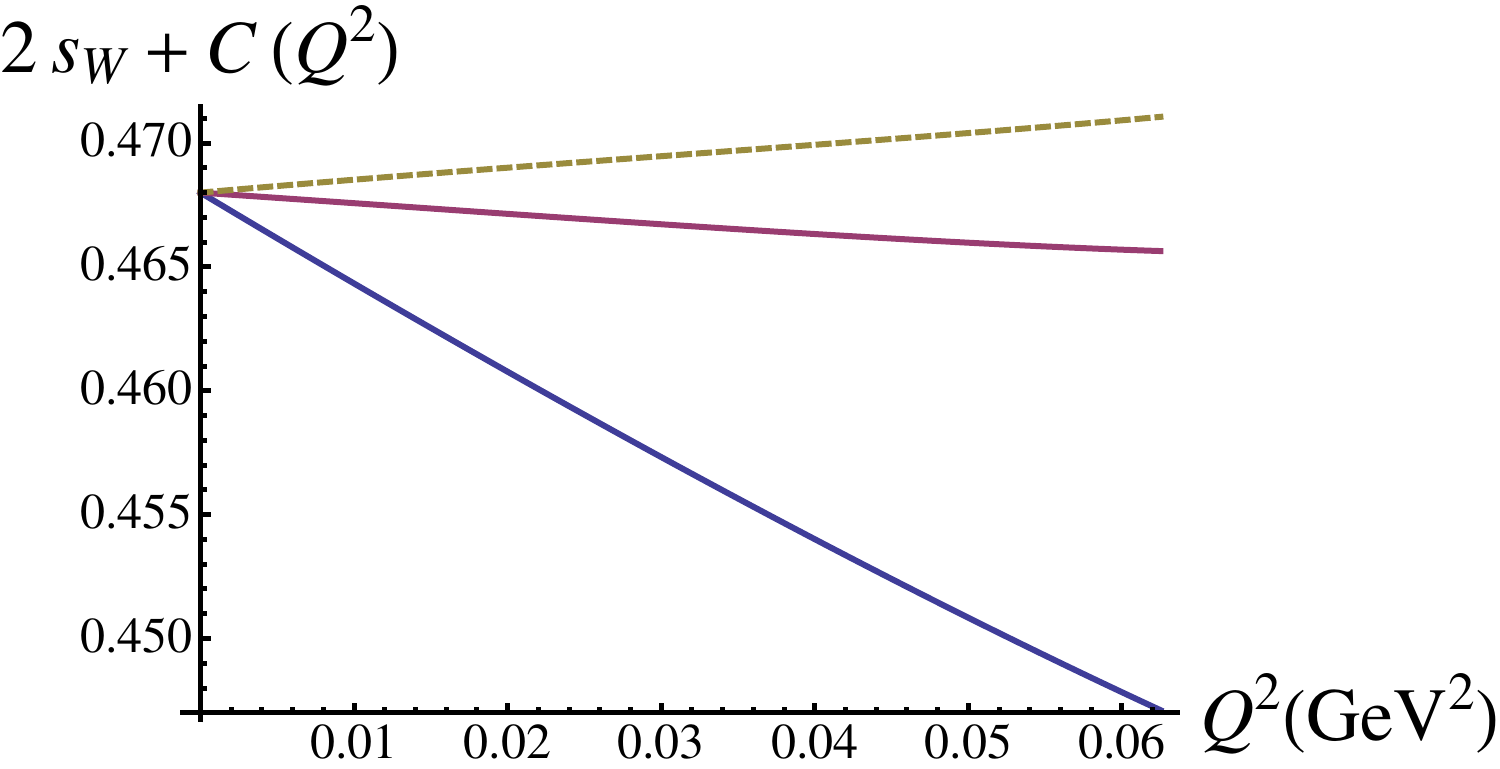}
\caption{(color online) $2 s_W+C(Q^2)$. The two solid curves include the effects of nuclear isospin violation, the average of nucleon CSB, and the upper and lower limits of the strangeness contribution.  The dashed curve is obtained by setting the strangeness contribution to 0. 
The use of any and each of the parameter sets discussed in this paper would lead to a straight line on this figure.
}\label{final}\end{figure}

Finally we plot the quantity $2s_W+C(Q^2)$ which gives via \eq{cdef} the PV asymmetry in units of
$G_FQ^2\over 2\pi\alpha\sqrt{2}$, see Fig.~\ref{final}.  
The two solid curves result from using the previously stated~\cite{Gonzalez-Jimenez:2014bia} upper and lower limits on 
$\rho_s$. A third dashed curve sets the strangeness contribution to zero ($\rho_s=0$).
Recent work relates the  strangeness contribution to deep inelastic scattering to that to proton electromagnetic form factors~\cite{us} through the use of light-front models, and  $\rho_s$ is limited to values about 10 times smaller than in~\cite{Gonzalez-Jimenez:2014bia} are obtained. If these models are valid, the dashed curve (with dominant contribution arising from nuclear isospin violation) would be the best prediction.

We summarize. The parity-violating elastic-$^{12}$C scattering asymmetry $\cal A$   at very low values of $Q^2$ is dominated by the size of the weak mixing angle, $s_W$. 
All of the corrections to that value  are linear in $Q^2$,
for the relevant range of  $0\le Q^2\le 0.0625$ GeV$^2$.  
 The CSB effects on nucleon electromagnetic form factors are at least an order of magnitude smaller than the 
 contributions expected from nuclear isospin breaking, which themselves are about $ 10^{-3}$ of the weak nuclear charge at $Q^2=0.01$ GeV$^2$.  The effects of nucleon strangeness are uncertain, but are linear with $Q^2$ in the relevant kinematic range.
 A measurement of the weak mixing angle  to the desired relative accuracy  of 0.3\% in the weak charge of $^{12}$C would require the ability to determine the  slope of $C(Q^2)$ to that accuracy to distinguish
a deviation from the standard model  from an effect of the correction term. This requires a measurement at more than one value of $Q^2$.


\begin{acknowledgments}
   This material is based upon work supported by the U.S. Department of Energy Office of Science, Office of Basic Energy Sciences program under Award Number DE-FG02-97ER-41014. I thank M. Wagman for technical help and
   for useful discussions.  I thank K. Kumar for suggesting this investigation and 
   Shi Jin for providing tables of  the proton and neutron densities of $^{12}C$ of Ref.~\cite{ab}.
   \end{acknowledgments}


\begin{thebibliography}{00}

\bibitem{Fei75} G. Feinberg, Phys. Rev. D 12 (1975) 3575.

\bibitem{Wal77} J. D. Walecka, Nucl. Phys. A 258 (1977) 349.

\bibitem{Don79} T. W. Donnelly and R. D. Peccei, Phys. Rep. 50 (1979) 1.

\bibitem{Mus94} M. J. Musolf, T. W. Donnelly, J. Dubach, S. J. Pollock,
S. Kowalski and E. J. Beise,  Phys. Rep. 239 (1994) 1.

\bibitem{Gerz:2013ema} 
  K.~Gerz, D.~Becker, S.~Baunack, K.~S.~Kumar and F.~E.~Maas,
  AIP Conf.\ Proc.\  {\bf 1563}, 86 (2013).

\bibitem{Moreno:2008ve} 
  O.~Moreno, P.~Sarriguren, E.~Moya de Guerra, J.~M.~Udias, T.~W.~Donnelly and I.~Sick,
  Nucl.\ Phys.\ A {\bf 828}, 306 (2009)
  \bibitem{GonzalezJimenez:2011fq} 
  R.~Gonzalez-Jimenez, J.~A.~Caballero and T.~W.~Donnelly,
  Phys.\ Rept.\  {\bf 524}, 1 (2013). 
  
\bibitem{Gonzalez-Jimenez:2014bia} 
  R.~Gonz‡lez-JimŽnez, J.~A.~Caballero and T.~W.~Donnelly,
  Phys.\ Rev.\ D {\bf 90}, 033002 (2014)
  
  \bibitem{Kelly:2004hm} 
  J.~J.~Kelly,
  Phys.\ Rev.\ C {\bf 70}, 068202 (2004).
  \bibitem{Wagman:2014nfa} 
  M.~Wagman and G.~A.~Miller,
  Phys.\ Rev.\ C {\bf 89}, 065206 (2014)  
\bibitem{Kubis:2006cy} 
  B.~Kubis and R.~Lewis,
  Phys.\ Rev.\ C {\bf 74}, 015204 (2006)
\bibitem{Bentz:2009yy} 
  W.~Bentz, I.~C.~Cloet, J.~T.~Londergan and A.~W.~Thomas,
  Phys.\ Lett.\ B {\bf 693}, 462 (2010) 
  \bibitem{us} T. J. Hobbs, M. Alberg,  G. A. Miller, ``DIS implications for nucleon strangeness", arXiv:1412.4871
  to be submitted to Phys. Rev. C.
\bibitem{ab} A. Bulgac, M. M. Forbes and S. Jin, ``A Nuclear Energy Density Functional",
to be published and private communication. 

\bibitem{AUERBACH:1972}
  N.~Auerbach, J.~Hufner,  A.~K.~Kerman and C.~M.~Shakin,
  Rev.\ Mod.\ Phys.\  {\bf 44}, 48 (1972).

\bibitem{HM79}
  E. M. Henley and G. A. Miller, in {\it Mesons in Nuclei}, edited by M. Rho and D. H. Wilkinson (North-Holland, Amsterdam, 1979), p. 405

\bibitem{Miller:1990iz} 
  G.~A.~Miller, B.~M.~K.~Nefkens and I.~Slaus,
  Phys.\ Rept.\  {\bf 194}, 1 (1990).
  G. A. Miller, B. M. K. Nefkens, and I. ?laus, Phys. Rep. 194, 1 (1990).
\bibitem{Dumbrajs:1983jd}
  O.~Dumbrajs, R.~Koch, H.~Pilkuhn, G.~C.~Oades, H.~Behrens, J.~J.~De Swart and P.~Kroll,
  Nucl.\ Phys.\ B {\bf 216}, 277 (1983).

\bibitem{Ericson:1988gk} 
  T.~E.~O.~Ericson and W.~Weise,
\textit{  Pions And Nuclei}, The  International  Series of Monographs on  Physics, Vol. 74,
  Oxford University Press, New York 1988

\end{thebibliography}
\end{document}